# Distributed Training of Deep Learning Models: A Taxonomic Perspective

Matthias Langer†, *Member, IEEE,* Zhen He‡, Wenny Rahayu‡ and Yanbo Xue†, *Member, IEEE*

**Abstract**—Distributed deep learning systems (DDLS) train deep neural network models by utilizing the distributed resources of a cluster. Developers of DDLS are required to make many decisions to process their particular workloads in their chosen environment efficiently. The advent of GPU-based deep learning, the ever-increasing size of datasets and deep neural network models, in combination with the bandwidth constraints that exist in cluster environments require developers of DDLS to be innovative in order to train high quality models quickly. Comparing DDLS side-by-side is difficult due to their extensive feature lists and architectural deviations. We aim to shine some light on the fundamental principles that are at work when training deep neural networks in a cluster of independent machines by analyzing the general properties associated with training deep learning models and how such workloads can be distributed in a cluster to achieve collaborative model training. Thereby we provide an overview of the different techniques that are used by contemporary DDLS and discuss their influence and implications on the training process. To conceptualize and compare DDLS, we group different techniques into categories, thus establishing a *taxonomy of distributed deep learning systems*.

✦

## 1 INTRODUCTION

THE current interest in deep learning was significantly spurred by breakthrough research results, which made it possible to build models to solve complex tasks, such as computer vision, speech recognition, natural language processing, etc [1], [2], [3], [4]. The massive parallel processing power of graphics processing units (GPUs) has been largely responsible for the recent successes in training deep learning models [5]. Hardware manufacturers continue their efforts to improve GPUs to speed up deep learning. However, as the hardware speed increases, so does the demand for deep learning models [1], [6]. In particular, the disruptive trend towards big data has led to an explosion in the size and availability of training datasets for machine learning tasks.

To be competitive, increasingly larger and more complex deep learning models are necessary [5]. Training such models on large datasets to convergence can easily take weeks or even months on a single GPU [4], [7]. A simple, but effective remedy to this problem is to utilize multiple GPUs to speed up training. Scale-up approaches rely on tight hardware integration to improve the data throughput. These solutions are effective, but costly. Furthermore, technological and economic constraints impose tight limitations on *scaling up* [8]. In contrast, distributed deep learning systems (DDLS) aim at *scaling out* to train large models using the combined resources of clusters of independent machines.

However, data transmissions across machine boundaries are much slower than performing equivalent transactions in a scaled-up system [8]. To train a large deep learning model using iterative stochastic gradient descent (SGD), the DDLS is required to frequently synchronize states and exchange intermediate representations of the model [9], [10]. As a cluster grows, so does the number of otherwise independent machines that have to be coordinated. At the same time, GPU-based deep learning greatly increases computation speeds [3], which often makes the network communication channel the primary bottleneck in distributed setups [11]. High-performance networking hardware can alleviate this problem, but is significantly more costly. Therefore, many DDLS proposed in the literature often take different approaches to realize distributed model training. This survey provides an overview of the current state-of-the-art in DDLS to aid researchers and practitioners in choosing appropriate techniques to harness the power of distributed computing infrastructures when training deep learning models.

Our contributions can be summarized as follows:

- We thoroughly analyze various distributed deep learning approaches, giving readers insight into the motivations and concepts behind different design choices and their influence on model training.
- These design choices form a taxonomy of the existing literature, with specific emphases on the implications from scaling out using a cluster of machines.
- Some DDLS differ significantly; others only subtly. We discuss a variety of major works in this field (scientific and commercial alike), show how they relate and use our taxonomy to categorize them.

The remainder of this survey is organized as follows. In Section 2, we differentiate our taxonomy-based approach from similar works. In Section 3, we discuss the major components at work in a DDLS and thereby develop a taxonomy for DDLS. In Section 4, we discuss various existing DDLS, explain how they relate to each other via our taxonomy and provide some pointers regarding how to choose a suitable technique. In Section 5, we summarize our work and point out potential directions for future research.

## 2 RELATED WORKS

There has been a number of different existing works that compare DDLS. Schmidhuber [2] provides a chronological overview of the recent developments in deep learning including certain advances in DDLS. Compared with this approach, our taxonomy is solely focused on the distributed problem domain and is organized by topic rather than chronologically. Zhang et al. [12] show how particular implementation decisions influence the training and network bandwidth usage in four distributed systems. Our approach is more principled. We identify the concepts that underpin the different methods of training commonly used in distributed deep learning and eventually lead to deviating implementation choices. Hence, we aim for a wider scope.

Ben-Nun et al. [13] recently published a tutorial/survey of parallel and distributed algorithms for deep learning that starts with a tutorial on general concepts like supervised learning, backpropagation and model architectures, before moving onto parallel and distributed training related topics (hyper-parameter and architecture search, etc.). We assume that the reader knows the fundamentals of deep learning and is interested in the best ways to train deep learning models in a cluster environment. Thus, rather than focusing on breadth, our survey offers more in-depth information on methods for distributed training. Our taxonomy-based approach aims to structure and compartmentalize the problem domain, provides a thorough analysis of the different design choices for training in a DDLS and explains the intuitions that underpin them (e. g. we expose, dissect and differentiate the inner workings of centralized and decentralized optimization under different scheduling regimes in Section 3.3). Furthermore, we also discuss topics and spotlight recent findings that are not or only tangentially covered by [13], but of high relevance for practitioners (e. g. the influence of staleness and its mitigation in asynchronous systems in Section 3.3.3).

## 3 TAXONOMY

Our aim is to provide a systematic overview over a wide range of principles and techniques used in DDLS. To achieve this, we create a taxonomy by assorting the fundamental characteristics that have a major impact on how DDLS operate. Understanding the intuitions and principles that underpin these characteristics allows actual DDLS to be interpreted as specializations of more generic concepts. We discuss these concepts with the limitations of distributed environments in mind. Applying our taxonomy (cf. Section 4) grants an unobstructed view of existing works in this field and enables comparisons to be made based on fundamental design decisions.

Furthermore, distributed deep learning has emerged rather recently. Existing literature in this area (cf. [2], [12], [13]) uses a variety of terminologies and divisions of the problem domain. Our taxonomy is an attempt to unify the terminology to enable certain phenomena to be precisely ascribed to characteristic properties or design choices.

Our taxonomy is split into 4 sections: 1) model vs. data parallelism; 2) centralized vs. decentralized optimization; 3) synchronous vs. asynchronous scheduling; and 4) the communication pattern used for exchanging parameters.

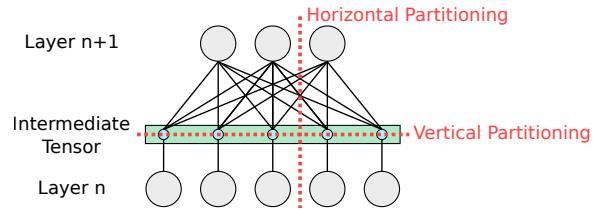

Fig. 1. Model parallelism can be achieved either through horizontal or vertical partitioning of the model.

### 3.1 Model vs. Data Parallelism

Model and data parallelism are two strategies for scaling out large deep learning workloads. The fundamental difference between both approaches is that model parallelism attempts to map the model execution steps onto the cluster hardware, while data-parallel approaches tackle collaborative model training as a concurrency/synchronization problem. Being orthogonal concepts, model and data parallelism can be applied simultaneously to achieve hybrid parallelism [14].

#### 3.1.1 Model Parallelism (MP)

In model parallelism the model is split into partitions, which are then processed in separate machines. To perform inference or train the model, signals have to be transmitted between them, so that each partition can be evaluated.

Training a deep learning model with SGD requires temporarily remembering the intermediate layer outputs observed during inference. Often, their size vastly exceeds that of the model [15]. Thus, partitioning the model and splitting the workload across machines can also help if the memory of any individual machine is not sufficient to store all model parameters [16].

Model partitioning can be conducted either by applying splits between neural network layers (=*vertical partitioning*) or by splitting the layers (=*horizontal partitioning*), as depicted in Fig. 1.

*Vertical partitioning* can be applied to any deep learning model because the layers themselves are unaffected. The surrounding logic (i. e. the DDLS) is responsible for transporting the intermediate tensor outputs of *Layer n* to the machine that executes *Layer n + 1*. During backpropagation, the error derivatives w.r.t. the loss function are passed through these layers (i. e. from machine to machine) in the reverse order. Although deep learning models can be split between any two layers, transitions between partitions that are located on different physical machines can be costly. The ideal partitioning depends on many factors (incl. the capabilities of the cluster hardware, shape of intermediate tensors, specific data flow during inference and backpropagation, etc.). Modern model-parallelism-capable DDLS like TensorFlow [17] employ heuristics and adaptive algorithms to determine efficient vertical partitioning schemes.

In *horizontal partitioning*, the layers themselves are partitioned. Hence, different parts of each training sample are processed in parallel using multiple devices. Thereby, horizontal partitioning often leads to a subset of neuron connections crossing partition boundaries. Efficiency hinges on finding splits that minimize the number of signals that have to transition machine boundaries. However, reorganizing and dispatching the individual layer outputs, such that they are consumable by each destination layer partition and vice-versa is complex and requires the DDLS to have detailed knowledge of the internal workings of the partitioned layers, which makes implementing this type of model partitioning across machine boundaries tedious in practice. Usually, horizontal partitioning is considered as a last resort if there is no other way to fit a layer into the memory of any single machine [9], or if the model contains large distinct sections with limited connectivity (e. g. non-convolutional locally receptive fields [18]).

Regardless of which partitioning strategy is used, the slowest route through the model determines the time required to perform inference and backpropagation. Whether an actual model training task can benefit from mapping the computation steps of the model onto the cluster hardware using model partitioning is highly situational [17]. Partitioning a model, such that the overhead is minimal and there are no bottlenecks, requires sophisticated algorithms in practice [17], [19]. Particular properties of the cluster configuration and any adjustment of the mini-batch size, model or computation graph (e. g. because it is data dependent) changes the optimal layout. Therefore, recent years have witnessed a shift away from model and towards data parallelism.

Pipelining signals through the partitioned model can help to better utilize the cluster hardware and increase the overall data throughput. However, during training each machine can only update its model parameters once all downstream computation steps for a mini-batch have been completed. Two possible ways to implement pipelining during training are: 1) Splitting the input mini-batch further, pipelining the fragments and accumulating the per-parameter gradients for the entire mini-batch, which are applied at the end. This does not solve the underlying problem, but, assuming the incurred overheads are low, results in a higher average GPU utilization [20]. 2) If enough resources are available to cache intermediate states for multiple mini-batches, pipelining entire mini-batches is also possible. However, this method leads to gradients being computed from stale parameters, a problem that is also frequently witnessed in certain data-parallel systems (cf. Section 3.3.3).

### 3.1.2 Data Parallelism (DP)

The basic idea underpinning data parallelism is to increase the overall sample throughput rate by replicating the model onto multiple machines, where backpropagation can be performed in parallel, to gather more information about the loss function faster. Conceptually, data parallelism is accomplished as follows. First, each cluster node downloads the current model. Then, each node performs backpropagation using its assignment of data in parallel. Finally, the respective results are aggregated and integrated to form a new model [21].

This is permissible because most transformations applied to a specific training sample in deep neural networks do not involve data from other samples[1]. Thus, the sum of per-parameter gradients computed using subsets $(\mathbf{x}^0, \cdots, \mathbf{x}^n)$ of a mini-batch $(\mathbf{x})$ matches the per-parameter gradients for the entire input batch (i. e. $\frac{\partial \mathcal{L}(\mathbf{x};w)}{\partial w} \approx \frac{\partial \mathcal{L}(\mathbf{x}^0;w)}{\partial w} + ... + \frac{\partial \mathcal{L}(\mathbf{x}^n;w)}{\partial w}$). Hence, assuming the mini-batch size is 64 samples in a cluster with two identical machines, then each machine may process 32 samples in parallel without requiring cross-machine communication. Typically, halving the number of training samples also halves the number of computations and the size of intermediate tensors, which speeds up backpropagation and is also helpful when working with large models.

---

[1]. This is not true if e. g. batch normalization [22] is used. However, if the individual mini-batch subsets are large enough, their statistics should approximate those of the entire mini-batch. Due to the intended normalization effect, minor variations of the statistics can even be desirable [8], which makes DP also applicable with such models.

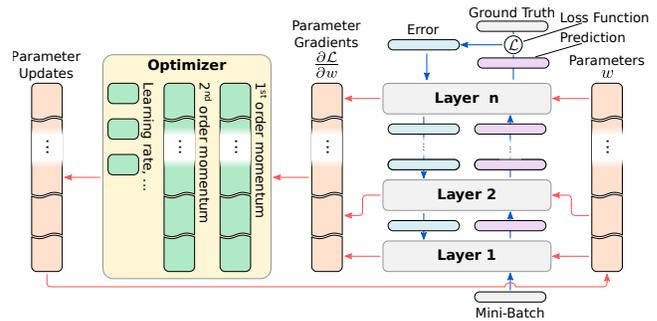

Fig. 2. Distinct data flow cycles in deep learning models during training (▲ = gradient computation cycle; ▲ = model update / optimization cycle).

Because either per-parameter gradients or the model parameters have to be transferred between machines, the relation of the model size and network bandwidth is key to whether data parallelism can accelerate training. The smaller a model is in comparison with its computational complexity, the easier it becomes to implement data parallelism [15]. For large models, bandwidth-related issues can quickly limit scalability [11]. However, as we will show in further parts of this taxonomy, data-parallel DDLS can apply various tricks to reduce the impact of bandwidth limitations.

Conceptually, further scaling out data-parallel systems just requires replicating the model code to another machine and assigning it to different mini-batch of data. This is in stark contrast to model-parallel approaches, where adding or removing a machine typically requires reevaluating the entire partitioning schema, which is significantly more challenging. Model parallelism remains relevant for in-node scaling if the cluster nodes are equipped with multiple GPUs [17], [19], [23]. With respect to the distributed domain, most recently developed DDLS focus primarily (occasionally solely; cf. [11], [24], [25], [26], [27]) on data parallelism. Therefore, the remainder of our taxonomy is predominantly focused on techniques used in data-parallel DDLS.

### 3.2 Centralized vs. Decentralized Optimization

In Fig. 2, we detail the data flow while training a deep learning model. The training procedure can be split into two distinct cycles. The blue process (▲) computes per-parameter gradients based on the current model parameters by applying backpropagation on mini-batches drawn from the training data. The optimization cycle (▲) consumes these gradients to determine model parameter updates. While this might appear as a bidirectional dependency, it is important to note that given a set of parameters, the model cycle can make many predictions to refine gradient outputs, while the optimizer needs updated gradients to progress.

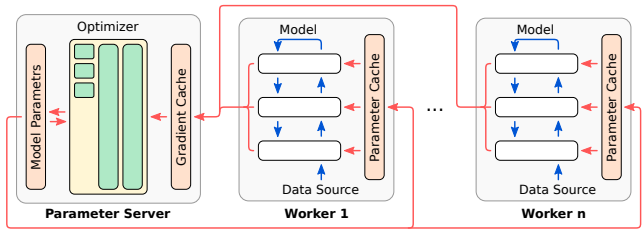

Fig. 3. Data flow in a cluster that implements centralized optimization. Responsibilities are separated. Workers evaluate the model to generate gradients. The parameter server consumes them to update the model.

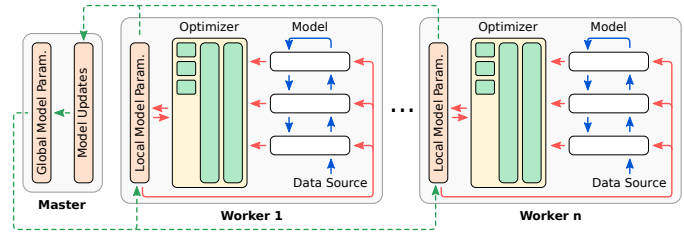

Fig. 4. Data flow in a cluster that implements decentralized optimization. The master node forms the next global model state by combining local model replicas (◄--) that were trained in isolation by the workers (⇒).

There are two major ways to map this execution model onto a cluster of independent machines: 1) *Centralized optimization*: The optimization cycle is executed in a central machine, while the gradient computation code is replicated onto the remaining cluster nodes. 2) *Decentralized optimization*: Both cycles are replicated in each cluster node and some form of synchronization is realized that allows the distinct optimizers to act cooperatively.

### 3.2.1 Centralized Optimization

In DDLS that implement centralized optimization, only a single optimizer instance (often called *parameter server*) is responsible for updating a specific model parameter. Parameter servers depend on the gradients computed by cluster nodes that perform backpropagation (*workers*). Fig. 3 illustrates the data flow during training in such a system. Note that the terms parameter server and worker refer to software processes, rather than actual machines. For simplicity, we assume for now that each process runs on a different machine. We will discuss how these roles can be organized to improve efficiency in Section 3.4.

Centralized optimization allows the expensive task of computing per-parameter gradients to be distributed across the cluster machines and elegantly handles updating the model by pooling all communication at the parameter server. Thereby, per-parameter gradients for large amounts of training samples can be computed quickly. Depending on whether computations across workers are scheduled synchronously or asynchronously, this can have different effects on the optimization (see Section 3.3).

Because a distinct parameter server is the only actor with write access for a specific model parameter, its state always reflects the current training progress. This greatly simplifies data handling, but leads to a producer-consumer relationship with the workers. Each worker has to re-download the model frequently to ensure that its produced gradients are relevant (cf. Section 3.3). For large clusters, this frequent need for communication focused at the same network endpoints can quickly become a bottleneck [11], [26]. Therefore, most centralized-optimization-based DDLS implement communication patterns where the parameter server role is distributed (cf. Section 3.4.1).

### 3.2.2 Decentralized Optimization

DDLS that rely on decentralized optimization treat their workers as a swarm, in which each worker independently probes the loss function to find gradient descent trajectories to minima that have good generalization properties [25]. Thus, instead of mapping the optimization cycle onto the cluster as in centralized DDLS, decentralized systems perform model training separately in each worker. To arrive at a *better* joint model, some form of arbitration is necessary to bring the different views into alignment [8]. Decentralized optimization cannot be used if there are workers that fail in fulfilling the memory requirements for replicating both, the gradient computation and optimization cycle.

Fig. 4 depicts the data flow in a decentralized system. Each worker represents an independent learner that repeatedly observes the loss function and adjusts its local model parameters to further decrease the loss. To achieve collaborative training, the workers have to exchange model parameters with each other. In this example, we assume the existence of a dedicated master node, which processes the individual parameter adjustments suggested by the workers and comes up with a new global model state that is then shared with them (◄--). Because model training and parameter exchange are decoupled, the master node's state is only loosely related to the different worker models. This property is characteristic for decentralized systems. Since the workers can make training progress without any communication, the network I/O bandwidth demands of decentralized systems are usually lower than those of their centralized counterparts [11], [25].

The key idea behind decentralized optimization is that multiple independent entities concurrently try to solve a similar but not exactly the same problem. Because the loss function in deep learning is usually non-trivial [28], independent numeric optimizers that observe the loss function at different locations eventually find different descent trails more appealing and converge towards different local minima. Hence, over time, the workers diverge and eventually arrive at incompatible models (i.e. models that cannot be merged without destroying the accumulated information). Therefore, DDLS that rely on decentralized optimization have to take measures to limit divergence [10]. In Fig. 5, we visualize the loss function landscape of a very simple model to demonstrate how decentralized optimization works. The master and all workers start from the same model state $\tilde{w}_0$ on the yellow plateau (=high loss). Then, an *exploration phase* (↘) begins, where the workers iteratively evaluate the loss function using different mini-batches and independently update their local models. After a while, each worker has formed a unique viewpoint regarding what permutations of the model parameters work best. Further training would lead to divergence and eventually incompatible models [29]. To avoid this, the local optimizers are interrupted and an *exploitation phase* (↖) occurs, during which each worker shares its model updates with the master node, which in turn merges the updates to distill latent parameter adjustments that have worked better on average across the investigated portion of the training dataset [11]. The thereby revised new global state $\tilde{w}_1$ is then shared with the workers, which use it as the starting point of the next exploration phase. Thus, by repeatedly applying exploration and exploitation, decentralized optimization subsequently finds a minimum of the loss function that the majority of the workers consider to be favorable.

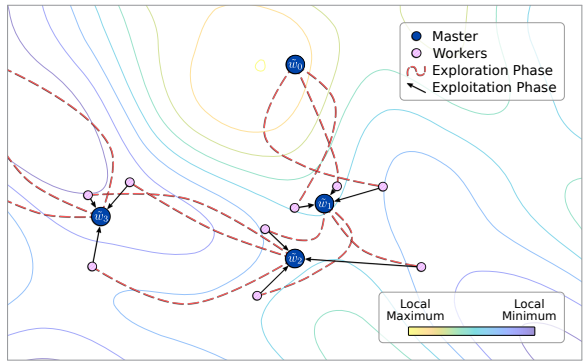

Fig. 5. Intuition behind decentralized optimization for an assumed loss function ($\mathcal{L}$). The objective is to find a local minimum (■) that generalizes well. Exploration phases, during which the workers probe $\mathcal{L}$ repeatedly and adjust their local representation of the model, always start from the most recent global state $\tilde{w}$ maintained by the master. Exploration phases are interleaved with exploitation phases, where the models are combined (e. g. through averaging [10]) to update the master's state.

Note how one worker in Fig. 5 is biased and gets trapped in the local minimum on the right. Because more workers eventually find the minimum on the left more attractive, the strolling worker is reset after the next exploitation phase. Although this worker has drifted away, its effort is not lost. By merging its divergent update proposals into the global model state, the overall convergence towards the left minimum is delayed, such that the surrounding loss landscape can be examined more thoroughly. Hence, in decentralized systems, each worker spends a significant portion of its resources on performing similar calculations that either disprove or confirm the findings of other workers. Both results can be used to further the development of the global model state. In case of disagreement, further adjustments of parameters are held back until progress has been made in other parameter space dimensions that are less disputed [30]. Analogously, agreement among workers can be utilized to accelerate the adjustment of undisputed parameters [11].

Having alternating exploration and exploitation phases that can be traded off against each other is common to all DDLS that rely on decentralized optimization. Therefore, decentralized DDLS tend to be more tolerant towards network bandwidth constraints. However, limits apply. Short exploration phases may not allow the workers to sufficiently probe the loss function to gather information, while long exploration phases lead to increased inconsistencies, which can hinder overall progress [10], [25]. To control the amount of exploration in relation to exploitation, decentralized optimization introduces additional hyper-parameters that need tuning to achieve fast convergence (see Section 3.3.2).

## 3.3 Synchronous vs. Asynchronous Scheduling

DDLS can also be distinguished into synchronous, asynchronous and bounded asynchronous systems. In bulk synchronous (or simply *synchronous*) systems, computations across all workers occur simultaneously. Global synchronization barriers ensure that individual worker nodes do not progress until the remaining workers have reached the same state. *Asynchronous* systems take a more relaxed approach to organizing collaborative training and avoid delaying the execution of a worker to accommodate other workers (i. e. the workers are allowed to operate at their own pace).

In other words, synchronous systems realize efficient collaborative training by avoiding deviations in progress between workers at the cost of a potential under-utilization of resources, while asynchronous systems favor a high hardware utilization to further training and regard deviations between workers as a manageable side effect that can—as we will show—even be advantageous in certain situations.

*Bounded asynchronous* systems represent a hybrid approach between these two archetypes. Fundamentally, they operate akin to centralized asynchronous systems, but enforce rules to accommodate workers progressing at different paces. Hence, the workers operate asynchronously with respect to each other, but only within certain *bounds*.

Due to the different motivations and effects of establishing synchronous or asynchronous modes of operation in conjunction with centralized and decentralized model training (see Section 3.2), we split their discussion accordingly.

### 3.3.1 Centralized Synchronous Systems

In centralized systems, model training is split between the workers (=gradient computation) and the parameter servers (=model update). If such a system exhibits a synchronous mode of operation, training cannot progress without a full parameter exchange between the parameter server and its workers, because the parameter server is dependent on the gradient input to update the model (cf. [15], [24], [27]). The workers, in turn, are dependent on the updated model in order to further investigate the loss function. Thus, in centralized synchronous DDLS, the cluster as a whole cyclically transitions between phases, during which all workers perform the same operation.

Fig. 6 shows the implementation of respectively the parameter server and worker programs of a simple centralized synchronous system. Each training cycle begins with the workers *downloading* new model parameters ($w$) from the parameter server. Then, they locally sample a training mini-batch ($\mathbf{x} \sim \mathcal{D}^i$) and *compute* per-parameter gradients ($g^i$). Thereafter follows a communication phase during which the workers *share* their gradients with the parameter server. The parameter server *aggregates* the gradients from all workers and injects the aggregate into an optimization algorithm to *update* the model.

Note that collaboration is only established through the aggregation of gradients. This is permissible as long as $\frac{\partial \mathcal{L}(\mathbf{x} \sim \mathcal{D}; w)}{\partial w} \approxeq \frac{\partial \mathcal{L}(\mathbf{x} \sim \mathcal{D}^1; w)}{\partial w} + \cdots + \frac{\partial \mathcal{L}(\mathbf{x} \sim \mathcal{D}^n; w)}{\partial w}$ is guaranteed (i. e. the aggregate of the per-parameter gradients derived from subsets of a large batch are approximately identical to those of the large batch itself), which is true for most operations frequently used in deep learning (cf. Section 3.1.2).

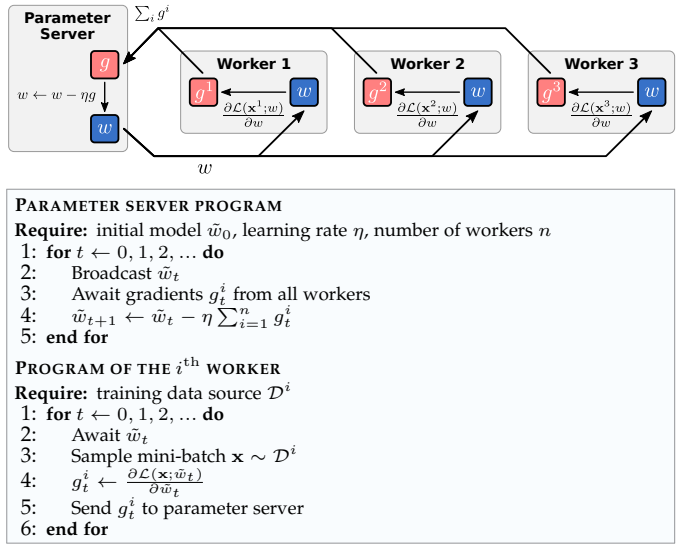

Fig. 6. Data flow in a centralized synchronous DDLS *(top)* and minimalist implementation of the parameter server and worker programs *(bottom)*. Note that global synchronization barriers are formed by the workers and parameter server awaiting each other's results.

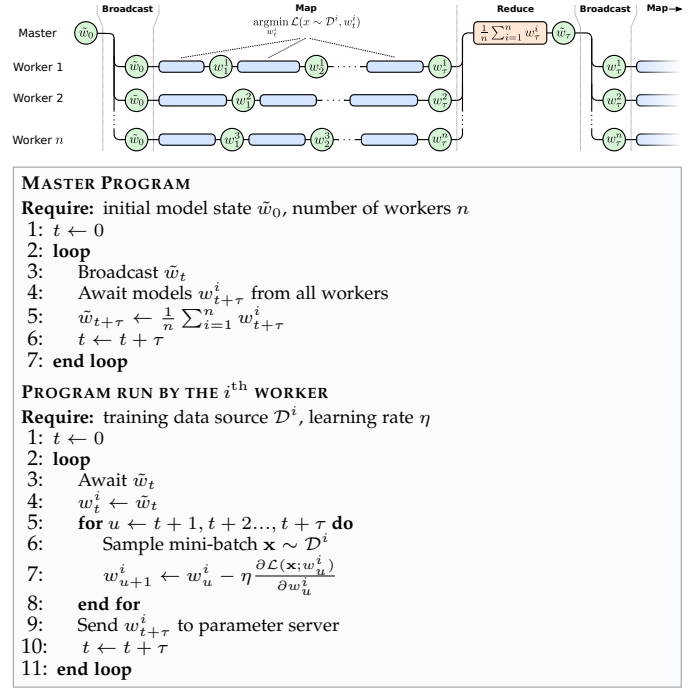

Fig. 7. Sequence diagram *(top)* and template implementation *(bottom)* of the decentralized synchronous system SparkNet [10].

Assuming appropriate and sufficient random sampling, larger mini-batches may represent the training distribution better [31]. As a result, the variance in the gradients between update steps decreases because the individual updates devised by the optimizer are based on a broader, more informed view of the training data. This can significantly speed up training when using accelerated gradient descent variants that scale updates depending on the stability of the gradient statistics (e. g. [32], [33]). However, note that the retained speedup from descent proposals derived from these less noisy gradients can diminish rapidly as the effective mini-batch size increases [8]. Furthermore, Keskar et al. [31] found that optimizing a model using mini-batches with a large coverage of the training distribution tend to get trapped in sharp minima basins of the loss function.

So far, we implied that the next training step can only be conducted once all workers have completed their assigned task and submitted gradients. In such configurations, a majority of cluster machines always has to wait for stragglers [34]. However, this is only mandatory if the training distribution is altered significantly without the contributions from certain workers. If the training set is 1) large enough, 2) reasonably well-balanced, and 3) sufficiently randomly distributed among the workers, it often does not matter whether minor portions of the training data are absent, such that this requirement can be relaxed. Thereby, it becomes possible to take shortcuts. For example, to avoid losing compute time Chilimbi et al. [23] proposed ending training epochs once 75% of all training samples have been processed, while Abadi et al. [17] suggested to generally over-provision by allocating more workers and ending each gradient aggregation phase once a quorum has been reached. Both approaches result in a significantly increased model update frequency, which typically more than compensates for the missing information from the delayed workers and the otherwise lost computational resources of fast workers.

### 3.3.2 Decentralized Synchronous Systems

Synchronous DDLS that rely on decentralized optimization independently conduct model training in each worker, and thus, do not exchange parameters to further model training, but rather to share the independent findings from each worker with the rest of the cluster to determine descent trajectories with good generalization properties (cf. Section 3.2). Thereby, they operate in phases separated by global synchronization barriers.

In Fig. 7, we respectively provide implementations of the master and worker programs and illustrate the sequence of events when training a model in the decentralized synchronous system SparkNet [10]. First, the initial model parameters ($\tilde{w}_0$) are distributed among the workers to initialize the local models ($w^i$). Then, the workers randomly sample mini-batches from their locally available partition of the training dataset, determine per-parameter gradients and adjust their model to minimize the loss function ($\mathcal{L}$). This process is repeated $\tau$ times, during which each worker independently trains its local model in isolation. This is the exploration phase. Due to the different properties of the mini-batches, each worker eventually arrives at a slightly better (w.r.t. $\mathcal{L}$), but different model. The master node acts as a synchronization conduit. After each exploration phase, the worker models ($w^i_\tau$) are merged to form a new joint model ($\tilde{w}_{t+\tau}$). This is the exploitation phase.

Note that unlike in centralized systems, gradients are not combined. The effective training batch size is identical to the mini-batch size used in each worker. Since optimizers only experience gradients with respect to the locally accessible portion of the training distribution, care must be taken when partitioning the training data to avoid situations where the loss function differs significantly between the workers.

As discussed in Section 3.2.2, decentralized DDLS have to balance exploration and exploitation. Keeping the local optimizers running for too long will result in reduced convergence performance or even setbacks if the worker models diverge[2] too far. In decentralized synchronous DDLS, the method of choice to address this problem is to limit the amount of independent exploration steps ($\tau$) [10], [25]. It has been shown that the best rate of convergence for a given model can typically be achieved if $\tau$ is rather small ($\tau \leq 10$; [8], [10]). However, since communication does not occur instantaneously, $\tau$ is not only a measure to control divergence, but also determines how much time should be spent on improving the local models versus synchronizing the states across machines. To make the best use of the cluster GPUs, $\tau$ should ideally be large, which often leads to sub-optimal convergence rates [11]. Hence, any choice of $\tau$ represents the dilemma of finding a balance between harnessing the benefits from having more computational resources and the need to limit divergence among workers. Studies by Zhang et al. [25] show that the influence of isolated learning on convergence can vary greatly depending on the cluster configuration and properties of the training task (e.g. the optimization algorithm). Because scaling out further is only sensible if the resources of additional machines are effectively utilized, practically motivated suggestions such as to aim for a 1:5 computation-to-computation ratio ($\approx$83.3% GPU utilization; [10]) may serve as a starting point for hyper-parameter search and to determine whether efficient decentralized optimization is possible at all using a certain configuration.

### 3.3.3 Centralized Asynchronous Systems

In centralized asynchronous DDLS (e. g. [14], [17], [19], [21], [35]), each worker acts alone and shares its gradients with the parameter server once a mini-batch has been processed. Instead of waiting for other workers to reach the same state, the parameter server eagerly injects received gradients into the optimization algorithm to train the model. Thus, each update of the global model is only based on the gradient input from a single worker. This is similar to the eager aggregation mechanisms discussed in Section 3.3.1. However, instead of discarding the results from all remaining workers and losing the invested computational resources, each worker is allowed to simply continue using its locally cached stale version of the model parameters.

We illustrate this idea in Fig. 8. The initiative in asynchronous systems remains with the workers, which approach the parameter server at their own pace to offer gradients, after which the global model is updated immediately, and request updated model parameters. In this way, each worker maintains a separate parameter exchange cycle with the parameter server. Because there is no interdependence between workers, situations where straggler nodes delay the execution of other workers cannot happen. For this system to work, choosing the results from one worker over

2. Instead of co-adapting by exploring adjacent closely related alternative models, the workers eventually split ways and develop fundamentally incompatible models that cannot be merged anymore.

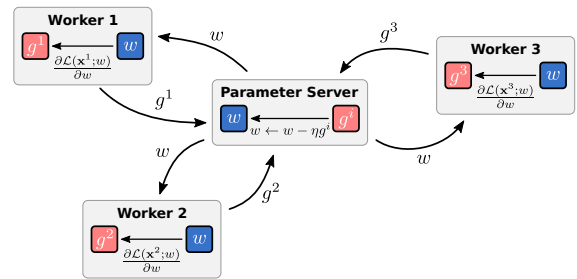

PARAMETER SERVER PROGRAM
**Require:** initial model state $\tilde{w}_0$, learning rate $\eta$
1: $\tilde{w} \leftarrow \tilde{w}_0$
2: Distribute $\tilde{w}$
3: **for** $t \leftarrow 0, 1, 2, \ldots$ **do**
4:     **if** received gradients ($g_{t-\delta}^i$) from worker $i$, with a delay of $\delta$ steps **then**
5:         $\tilde{w} \leftarrow \tilde{w} - \eta g_{t-\delta}^i$
6:         Send $\tilde{w}$ to worker $i$
7:     **end if**
8: **end for**

PROGRAM OF THE $i^{\text{th}}$ WORKER
**Require:** training data source $\mathcal{D}^i$
1: **for** $t^i \leftarrow 0, 1, 2, \ldots$ **do**
2:     Await of current parameter server model $\tilde{w}$
3:     $w^i \leftarrow \tilde{w}$
4:     Sample mini-batch $\mathbf{x} \sim \mathcal{D}^i$
5:     $g^i \leftarrow \frac{\partial \mathcal{L}(\mathbf{x};w^i)}{\partial w^i}$
6:     Send $g^i$ to parameter server
7: **end for**

Fig. 8. Data flow *(top)* and minimalist implementation of a centralized asynchronous DDLS *(bottom)*.

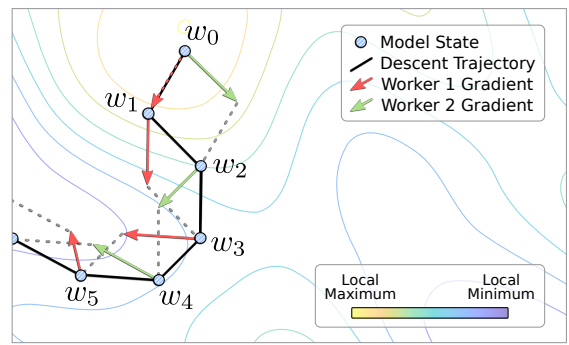

Fig. 9. Intuition for centralized asynchronous systems. We assume fair scheduling among workers in this example. Therefore, the worker that is the next to submit gradients to update the model state is always in possession of a stale model state.

another must not introduce a bias that significantly changes the shape of the loss function [36]. Thus, on average, the mini-batches sampled by each worker have to mimic the properties of the training distribution reasonably well.

A consequence of this mode of operation is that, at any point in time, only a single worker is in possession of the most recent version of the model. Other workers only possess stale variants that represent the state of the parameter server during their last interaction with it. Any gradients that they produce are relevant to the shape of the loss function around that stale model representation.

In Fig. 9, we illustrate how the gradient descent trajectory evolves in a small centralized asynchronous system. For the sake of clarity, we assume fair scheduling in this drawing. Both workers start from the same model state ($w_0$), but draw different mini-batches from the same distribution.

First, worker 1 transmits its gradients to the parameter server and the parameter server sends back the updated model $w_1$ to that worker, which in turn immediately uses it to compute new gradients. Meanwhile, worker 2 transmits gradients to the parameter server. These gradients were determined using the now outdated state $w_0$, while the resulting update step is applied to $w_1$ to form $w_2$. Based on $w_2$, worker 2 evaluates the loss function again. Hence, when worker 1, which was only aware of state $w_1$, returns to exchange updates with the parameter server, its gradients are also based on an outdated model state. Without visibility of the loss function landscape, the parameter server relies on the workers supplying relevant information and injects it into its optimization logic to update the model (dashed gray line; ---). The rationale here is that because these states are supposedly closely related to the current state of the parameter server, the gradient information submitted is still useful to improve the model (i.e. reduce the average loss).

In our two-worker-example, one worker always lags behind by at least one model update cycle. Model parameter updates derived from the submitted gradients do not necessarily represent the best decision to make. However, if the learning rate is small enough, the overall loss function landscape should not change too rapidly. Therefore, the increased amount of information about the loss function from having more computational resources helps to find descent trajectories that work better on average. One way to interpret this is that the individual workers explore side channels of the descent trajectory [8]. Depending on the quality of the side channel, the gradients submitted by a worker will add a bias to the model that either pushes it towards that side channel or away from it. For dimensions in which the overall descent trajectory across many side channels overlaps, parameter adjustments will be conducted faster. In disputed dimensions, progress will be slowed down due to increased variance of the gradients [32], [33].

Staleness has serious implications on model training [37]. In a cluster with multiple asynchronous workers, the next worker to exchange parameters is usually stale by some amount of update steps. The more workers are in the cluster, the more update steps the next worker lags behind on average and the less relevant the related gradients submitted are to the current training progress [38]. Applying delayed updates from stale workers can be interpreted as having an implicit momentum term in the loss function that scales with the number of workers [39]. For each deep learning task in a centralized asynchronous setting, there exists an upper bound, beyond which scaling out further will degrade the overall training performance due to overshooting effects from the increased implicit momentum. Mitliagkas et al. [39] show that compensating implicit momentum by extending the loss function with an explicit *negative momentum* term is possible and allows scaling out further under certain circumstances. However, different limits apply depending on the model and dataset. *Delay compensation* [40] is an alternative method to mitigate delay by adjusting the gradients in the parameter server before applying the optimizer based on the recent development of each model parameter since the submitting worker last downloaded the model. To achieve this, the parameter server has to remember all past model states that are still in use by any worker, which can become infeasible as the cluster grows larger.

However, note that regardless of the measures taken, any new gradients computed by the slowest of $n$ independent asynchronous workers pertain to the loss of a model that was adjusted at least $n-1$ times in the meantime. Thus, the information gain with respect to the current global model state still diminishes as the cluster grows larger.

### 3.3.4 Bounded Asynchronous Systems

The fair scheduling we implied in our analysis in Section 3.3.3 is undesirable in practice because the slowest machine would hold back faster machines, which is exactly the situation that asynchronous systems try to avoid. However, not enforcing any order when exchanging parameters carries some risk. Gradients from severely eclipsed workers can confuse the parameter server's optimizer, which can setback training or even destroy the model.

Aside from taking measures to ensure that the individual memory transactions are race-condition free to avoid overt information loss, early centralized asynchronous systems [21], [41] integrated gradients into the global model irrespective of their staleness. In contrast, most modern deep learning systems constrain the asynchronous operation of the workers to avoid the detrimental effects from severely stale updates by either prioritizing or delaying certain parameter exchange requests to establish order, enhance consistency and control staleness across the cluster. Each distributed system implements such coordination mechanisms slightly differently (cf. [16], [17], [19], [35]). In the following discussion, we will focus only on the most important methods.

To avoid compounding delays the parameter server typically places workers that indicated their readiness to upload gradients in a priority queue based on their staleness [17], [19], [42]. To protect against adverse influences from severe stragglers, some systems allow defining conditions that must be fulfilled before queued requests can be processed by the parameter server. These conditions typically take the form of either a value or delay bound.

*Value bounds* limit the change of model parameters that has not yet been shared with other workers. To track this, the parameter server maintains a copy of all versions of the model currently in use across the cluster (from the model that is currently known by the slowest worker $w_{t-\delta}$, to the most recent model $w_t$). $w_t - w_{t-\delta}$ is the amount of change in transit that is currently not known by the slowest worker. If a worker triggers an update that leads to a violation of some value bound (i.e. $\|w_t - w_{t-\delta}\|_\infty \geq \Delta_{\max}$), it is delayed until the value bound condition holds again. Since the magnitude of future model updates is largely unknown, choosing a reliable metric and limit for a value bound can be difficult and may require adjustment during training [38].

*Delay bounds* (e.g. the Stale Synchronous Parallel (SSP) criterion [35]): Each worker ($i$) maintains a separate clock ($t^i$). Whenever a worker submits gradients to the parameter server, $t^i$ is increased. If the clock of a worker differs from that of the slowest worker by more than $s$ steps, it is delayed until the slow worker has caught up. Thus, if a worker downloads the current global model it is ensured that this model includes all local updates and may also contain updates from other workers within a range of $[t^i - s, t^i + s - 1]$ update steps. Thereby, the amount of state change is scaled implicitly based on the recently observed shape of the loss function landscape [14]. It has been shown that choosing $s$ constant is sufficient to efficiently mitigate occasional delays in small- and medium-sized clusters [38]. Note that generally slower and severely delayed workers may still reduce the throughput of the entire cluster. Extensions of the basic algorithm like ESSP [16] attempt to address this problem through lowering the average staleness across workers by proactively distributing the current global model state.

Value and delay bounds are also useful for maintaining orderly behavior in setups where the parameter server role is split among multiple machines (see Section 3.4). For example, both approaches can easily be extended to implement intermediate caches [19], or to realize delay tolerant synchronization between concurrent parameter servers, which in turn permits the implementation of hierarchical or mesh-like cluster topologies that guarantee a certain level of consistency at all times [16]. An alternative to stalling progress in response to violating a bound is to *reject* updates from slow workers. However, we mind that this may introduce undesirable biases, which must be considered when distributing and/or sampling of the training dataset.

### 3.3.5 Decentralized Asynchronous Systems

Unlike in decentralized synchronous systems (see Section 3.3.2), where model training in all workers frequently ceases at a global synchronization barrier to re-parameterize the local models, the workers in decentralized asynchronous systems act independently and continue to explore the loss function based on a model that is detached from the master's current state. Consequently, the workers cannot replace their model parameters upon completing a parameter exchange with the master node. Instead, they have to merge the respective asynchronously gathered information. The method of choice (cf. [11], [25], [29], [30], [42], [43]) for combining master ($\tilde{w}$) and worker models ($w^i$) in such a setting is to apply linear interpolation as shown in Equation 1.

$$\begin{aligned} w^i &\leftarrow w^i - \alpha(w^i - \tilde{w}) \\ \tilde{w} &\leftarrow \tilde{w} + \beta(w^i - \tilde{w}) \end{aligned} \quad \text{for } \{\alpha, \beta\} \in \left[0, \frac{1}{2}\right] \text{ and } \alpha \geq \beta \quad (1)$$

Thus, upon exchanging parameters, the worker model is displaced towards the master model's state at a rate of $\alpha$ times their relative distance. The master model is displaced in the opposite direction at a rate of $\beta$. Note how this operation is equivalent to temporarily extending the loss function with the squared $\ell^2$-norm of the difference between both models (i.e. $\mathcal{L}(x^i, w^i) - \frac{\alpha}{2} \|w^i - \tilde{w}\|_2^2$).

In Fig. 10, we showcase example implementations of respectively the master and worker programs in the decentralized asynchronous system *Elastic Averaging SGD* (EASGD [30]). Since each worker ($i$) acts independently, it maintains a separate local clock ($t^i$) to measure the length of isolated learning. Once $\tau$ iterations have been completed by a worker, the master node's current model $\tilde{w}$ is downloaded and the penalization term $\delta^i$ is computed and applied to the local model. Then $\delta^i$ is transferred to the master node, which applies the inverse operation. $\delta^i$ has already incorporated $\alpha$. Hence, $\alpha = \beta$ in this particular implementation. There exists a symmetric force between each worker and the master node that equally attracts both models. Zhang et al. [30]

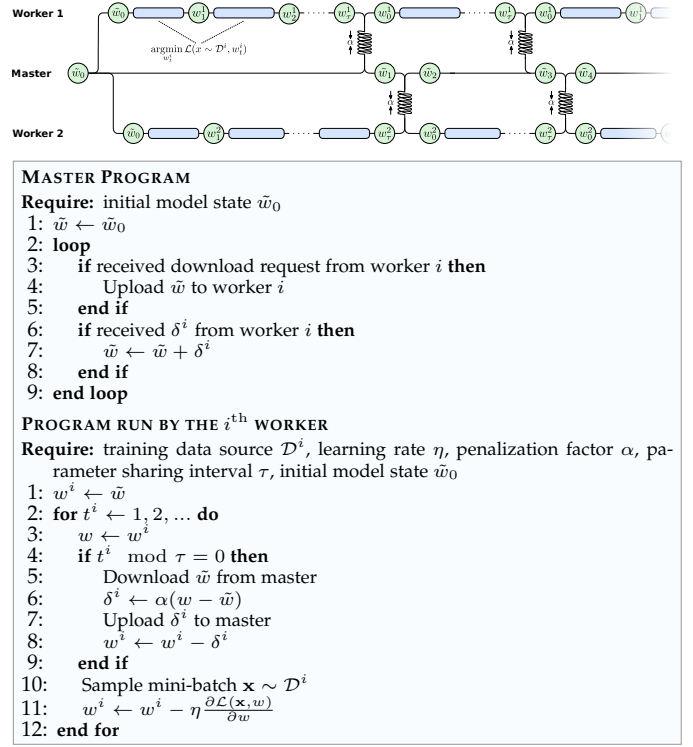

Fig. 10. Sequence of execution *(top)* and example implementation *(bottom)* of the decentralized asynchronous system EASGD [30].

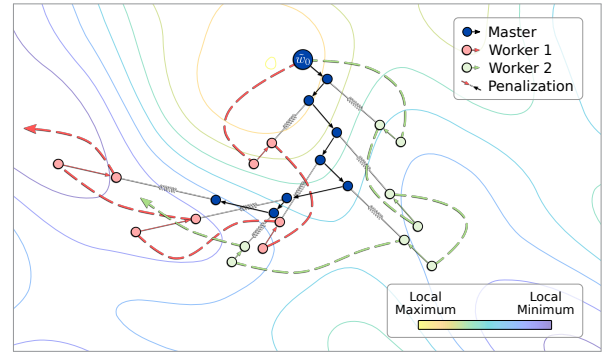

Fig. 11. Overall intuition for decentralized asynchronous systems. The workers detach from the master model and independently explore the loss function. The master's model is alternately attracted towards the current state of different workers and vice-versa—as if they were connected using springs. Through repeatedly applying this mechanism, the master and the workers subsequently drift towards model parameter adjustments that—on average—reduce the loss best (i. e. a minimum).

refer to this concept as elastic symmetry, which they assume to be crucial for stability. Fig. 11 illustrates the intuition for the developments occurring within the corresponding models. The individual models are evolving side-by-side in parallel. Once a worker has evaluated the loss function and applied the corresponding adjustments to its local model $\tau$ times, it contacts its master node to initiate a parameter exchange, which results in the displacement of both models towards each other. Note that there is no direct interaction between workers. Stability is maintained by the penalization coefficients ($\alpha$ and $\beta$) in combination with the length of isolated learning phases ($\tau$).

Because the workers independently initiate parameter exchanges, the communication demand with the master node scales roughly linear with the number of workers [30]. To avoid congestion induced delays due to network I/O bandwidth limitations at the master, $\tau$ must be scaled accordingly [11]. However, like in synchronous decentralized systems, long phases of isolated training can severely hamper convergence due to the increasing incompatibilities in the models (cf. Section 3.3.2).

The optimizer hyper-parameters, $\alpha$, $\beta$ and $\tau$, are interdependent and must be weighted carefully for each training task and cluster setup to constrain how far individual workers can diverge from the master and one another. However, their dynamics change throughout training, which can make decentralized asynchronous model training challenging in practice [25]. Some recent works in this area focus on improving this situation. For example, by actively adapting $\alpha$ based on the rate of convergence [42], or by adjusting the parameter exchange cycle in response to the momentary available I/O capacity and controlling the therefore differently progressing workers using a scalable penalization term [11].

### 3.4 Communication Pattern

So far, we implied that each function in the cluster is executed by a separate machine. However, workers, parameter servers and master nodes are software applications that can be run on the same machine, separate machines or spread across multiple machines. Distributing the worker role is essentially model parallelism, which we detailed in Section 3.1. In this section, we further our analysis and focus on communication patterns that data-parallel DDLS employ to organize the cluster and speedup parameter exchanges.

#### 3.4.1 Communication patterns in centralized systems

Regardless of the training method, having the parameter server role concentrated in a single machine greatly simplifies the overall system architecture because all model training is coordinated in a single software program. Besides, such systems are easy to configure, control and debug. However, this may make the parameter server a bottleneck.

For example, to perform a full parameter exchange in a cluster with $n$ workers, a dedicated single parameter server or master node needs to subsequently send and receive $n\|w\|$ parameters. In bulk-synchronous systems, updates of the joint model depend on the contributions from all workers. Parameter up- and downloads occur sequentially, implying a communication delay of at least $2nT_w + (n-1)R_w + U_w$ in theory, where $T_w$, $R_w$ and $U_w$ respectively denote the time required to *transmit*, *reduce* and *update* $\|w\|$ parameters (i.e. the model). However, actual DDLS are rarely implemented in this naïve way. Remember that the synchronization barriers impose cluster-wide phases during which only unidirectional parameter transfers occur from either the parameter server or master node to the workers, or vice versa, which enables the use of efficient collective communication primitives, such as binomial trees [15], or the scatter-reduce/broadcast algorithm [44], for which the lower bound communication delays in the same cluster setup are respectively only $2\lceil\log_2(n+1)\rceil T_w + \lceil\log_2(n)\rceil R_w + U_w$ and $(2 + 2\frac{n-1}{n})T_w + \frac{n-1}{n}R_w + U_w$. Note how this is in stark contrast to asynchronous systems, where the workers individually peer with the shared role *ad hoc*. There, assuming no congestion, the minimum communication delay is $2T_w + U_w$ per worker. Parameter exchange requests from individual workers can be overlapped if $n > 1$. Thus, in an ideal scenario, a full parameter exchange with all workers can be completed in $nT_w + U_w$.

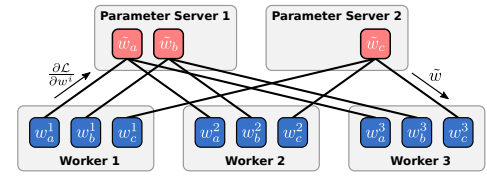

Fig. 12. Centralized system featuring multiple parameter servers. The model is partitioned into shards ($a$, $b$ and $c$). Each parameter server maintains a subset of these partitions.

If the parameter server is a bottleneck, it is highly desirable to distribute this role [21], such that communication is not focused on a single network endpoint (see Fig. 12). Most gradient-descent-based optimization algorithms can be executed independently for each model parameter, which permits almost arbitrary slicing. Of course, in practice this freedom is limited by the overheads incurred from peering with each additional endpoint to complete a parameter exchange [23]. This is especially true in asynchronous systems where congestion-free $k : n$ communication (i.e. between $k$ parameter servers and $n$ workers) is more difficult to realize because the workers operate largely at their own pace [25]. Additional limitations apply if training depends on hyper-parameter schedules that must be coordinated across parameter servers, or if reproducibility is desired [21].

Usually, the computational costs of the optimization algorithm are low in comparison to performing backpropagation. Therefore, a popular variant of the multi-parameter-server-approach is to migrate the parameter server role into the worker nodes (cf. [17], [19], [24], [27]), such that all nodes are workers, but also act as parameter servers (i.e. $k = n$). In Fig. 13 (left), we illustrate a cluster configuration where each worker is responsible for maintaining and updating $\frac{1}{n}$ of the global model parameters. This locally maintained model partition does not have to be exchanged via the network. Thus, the external communication demand of each node is reduced to $2\frac{n-1}{n}\|w\|$. In synchronous systems, communication delays can be further optimized through collective communication (e.g. via suitable scatter reduce/broadcast operations; $2\frac{n-1}{n}T_w + \frac{n-1}{n}R_w + \frac{U_w}{n}$ [44]). A neat property of this approach is that the workload and communication demand of all machines are identical, which can be beneficial in homogeneous cluster setups. However, any node-failure requires a complete reorganization of the cluster [12].

An alternative method that avoids this problem at the expense of memory and computational overhead ($U_w$ vs. $\frac{U_w}{n}$) is illustrated in Fig. 13 (right). Here, the entire parameter server function is implemented in each worker. The workers synchronously compute gradients, which are shared between machines using a collective *all-reduce* operation[3].

---

3. For example, via recursive doubling as depicted in Fig. 13 (lower bound delay = $\lceil\log_2(n)\rceil(T_w + R_w) + U_w$), or—more commonly—using the ring algorithm (lower bound delay = $2\frac{n-1}{n}T_w + \frac{n-1}{n}R_w + U_w$).

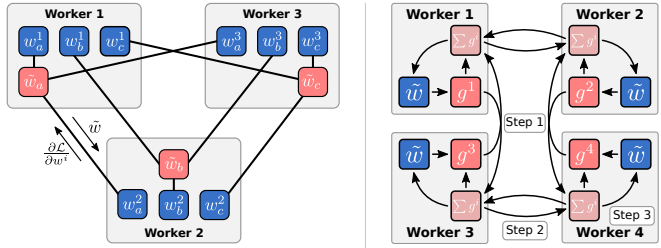

Fig. 13. Cluster setups for centralized systems where each worker implements parameter server functions. *Left:* Each worker is responsible for a portion of the model. *Right:* All workers update the entire model in sync using gradients collected from all workers via recursive doubling.

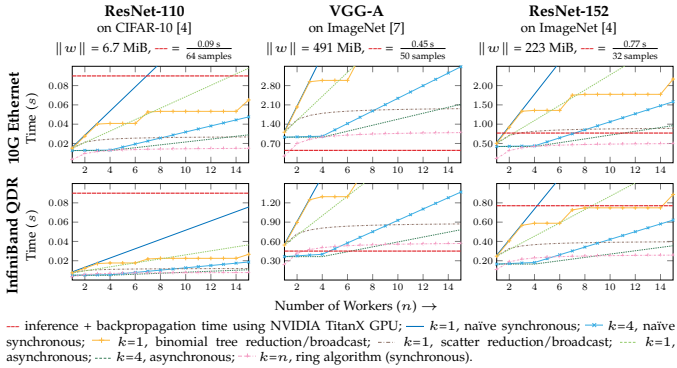

--- inference + backpropagation time using NVIDIA TitanX GPU; — $k$=1, naïve synchronous; — $k$=4, naïve synchronous; — $k$=1, binomial tree reduction/broadcast; -- $k$=1, scatter reduction/broadcast; ---- $k$=1, asynchronous; ---- $k$=4, asynchronous; -+- $k$=$n$, ring algorithm (synchronous).

Fig. 14. Lower bound communication delays when training various models in different cluster setups in Ethernet and InfiniBand environments, assuming $R_w \approx 10$ GiB/s and $U_w \approx 2$ GiB/s in an ideal scenario with no latency or competing I/O requests that need arbitration.

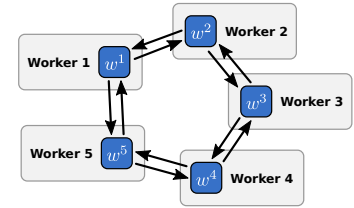

**Require:** training data source $\mathcal{D}^i$, initial model state $\tilde{w}_0$, learning rate $\eta$, neighbor nodes $N^{i-1}$ and $N^{i+1}$, merge weights $\alpha^{i-1,i,i+1}$
1: $w_0^i \leftarrow \tilde{w}_0$
2: **for** $t \leftarrow 0, 1, 2, \ldots$ **do**
3: $\quad$ Send $w_t^i$ to neighbor nodes $N^{i-1}, N^{i+1}$
4: $\quad$ Sample mini-batch $\mathbf{x} \sim \mathcal{D}^i$
5: $\quad g^i \leftarrow \frac{\partial \mathcal{L}(\mathbf{x}; w_t^i)}{\partial w_t^i}$
6: $\quad$ Await $w_t^{i-1}$ and $w_t^{i+1}$ from neighbor nodes
7: $\quad w_{t+1}^i \leftarrow \frac{1}{\sum_{j=i-1}^{i+1} \alpha^j} \left( \sum_{j=i-1}^{i+1} \alpha^j w_t^j \right) - \eta g^i$
8: **end for**

Fig. 15. Organization of workers *(top)* and corresponding worker program for the $i^{\text{th}}$ worker *(bottom)* in D-PSGD [26].

Each machine uses the thereby locally accumulated identical gradients to step an equally parameterized optimizer copy, which in turn applies exactly the same update. Note how this setup is not only robust to node failures, but also makes adding and removing nodes trivial.

In Fig. 14, we plot the lower bound delays for the aforementioned communication patterns when training image classification models in clusters of varying size and the corresponding gradient computation times (---). Keep in mind that these are the minimum delays assuming 100% of the bandwidth is available. In practice, concurrent access, processing speed deviations and network traffic by other processes induce additional delays. Generally speaking, communication delays in centralized synchronous systems should be minimized to retain a high GPU utilization, while for centralized asynchronous DDLS it is sufficient to keep the cumulative communication delay across all workers below the time it takes to do a full parameter exchange to avoid worker starvation [8]. This is not always possible. For some configurations shown in Fig. 14, the lower bound communication delay exceeds the time required for processing a reasonably sized mini-batch in a worker. As can be seen, using efficient communication patterns or multi parameter server cluster setups can dramatically change the picture, thus making data-parallel model training feasible. However, efficiency still hinges heavily on the model-size-to-computation-time ratio (cf. Section 3.1.2). Training a less complex model with many tunable parameters like VGG-A [7] using centralized methods can quickly exhaust even the capabilities of data-center-grade networking equipment.

Communication patterns where model parameters are frequently exchanged across machines without coordination can quickly become difficult to manage as a cluster grows larger. In particular in large asynchronous systems, conditions that need arbitration are unavoidable. Imposing bounds on the asynchronous processing is an effective measure to maintain order in such situations (cf. Section 3.3.4). But as the cluster grows larger, tight limits will eventually hold back more and more workers. Centralized DDLS that scale out to hundreds or more nodes often avoid I/O bottlenecks by structuring communication through introducing proxy servers that act as a bidirectional write-back cache [14], [19] (e.g. implemented as a value or delay bound; see Section 3.3.4) for distinct subsets of workers. However, note that having interim caches and/or proxy servers can introduce additional staleness (cf. Section 3.3.3).

### 3.4.2 Communication patterns in decentralized systems

Assuming isolated training phases of $\tau$ cycles, the communication demand of each decentralized worker per local compute step is only $\frac{1}{\tau}\|w\|$. In this way, decentralized systems typically maintain a higher computation hardware utilization, even with limited network bandwidth, which can make training large models possible in spite of bandwidth-constraints [11]. However, scaling out to larger cluster sizes may still result in the master node becoming a bottleneck. Of course, it is possible to split the master's role in decentralized systems to reduce communication costs like in centralized systems. However, because each machine is a self-contained independent trainer, decentralized DDLS have many options for organizing parameter exchanges.

For example, *D-PSGD* [26] implements a communication pattern that avoids the costly gathering and redistribution of the model across the entire cluster. Each worker only exchanges parameters with two partner nodes. Thereby, a ring-like structure is formed as illustrated in Fig. 15. After each training cycle, each worker sends its model parameters to its neighbors and also integrates the models it receives from them. Hence, each worker repeatedly influences and regularizes its neighbors and, thus, explores the nearby

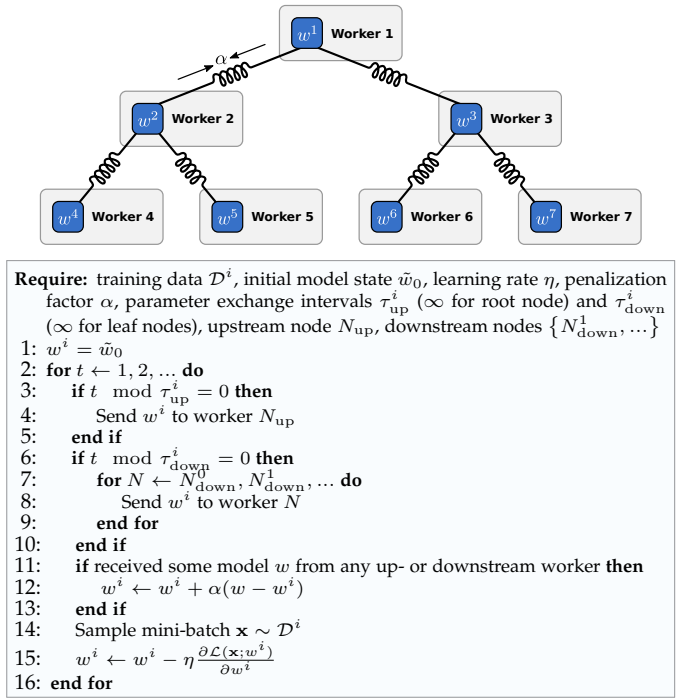

**Require:** training data $\mathcal{D}^i$, initial model state $\tilde{w}_0$, learning rate $\eta$, penalization factor $\alpha$, parameter exchange intervals $\tau_{up}^i$ ($\infty$ for root node) and $\tau_{down}^i$ ($\infty$ for leaf nodes), upstream node $N_{up}$, downstream nodes $\{N_{down}^1, ...\}$

1: $w^i = \tilde{w}_0$
2: **for** $t \leftarrow 1, 2, \ldots$ **do**
3:   **if** $t \mod \tau_{up}^i = 0$ **then**
4:     Send $w^i$ to worker $N_{up}$
5:   **end if**
6:   **if** $t \mod \tau_{down}^i = 0$ **then**
7:     **for** $N \leftarrow N_{down}^0, N_{down}^1, \ldots$ **do**
8:       Send $w^i$ to worker $N$
9:     **end for**
10:   **end if**
11:   **if** received some model $w$ from any up- or downstream worker **then**
12:     $w^i \leftarrow w^i + \alpha(w - w^i)$
13:   **end if**
14:   Sample mini-batch $\mathbf{x} \sim \mathcal{D}^i$
15:   $w^i \leftarrow w^i - \eta \frac{\partial \mathcal{L}(\mathbf{x}; w^i)}{\partial w^i}$
16: **end for**

Fig. 16. Relationship between workers *(top)* and program for the $i^{th}$ worker *(bottom)* in TreeEASGD [25].

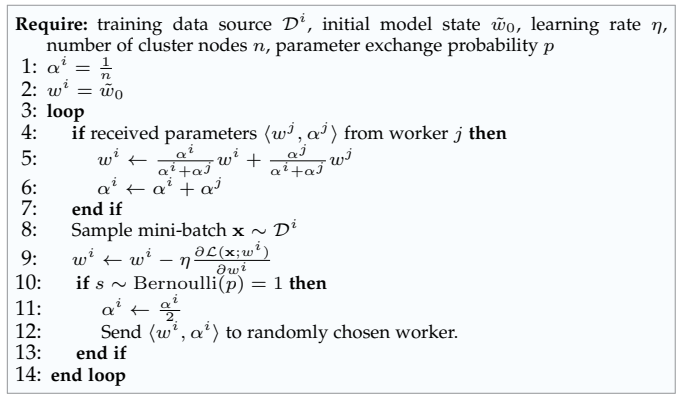

**Require:** training data source $\mathcal{D}^i$, initial model state $\tilde{w}_0$, learning rate $\eta$, number of cluster nodes $n$, parameter exchange probability $p$

1: $\alpha^i = \frac{1}{n}$
2: $w^i = \tilde{w}_0$
3: **loop**
4:   **if** received parameters $\langle w^j, \alpha^j \rangle$ from worker $j$ **then**
5:     $w^i \leftarrow \frac{\alpha^i}{\alpha^i + \alpha^j} w^i + \frac{\alpha^j}{\alpha^i + \alpha^j} w^j$
6:     $\alpha^i \leftarrow \alpha^i + \alpha^j$
7:   **end if**
8:   Sample mini-batch $\mathbf{x} \sim \mathcal{D}^i$
9:   $w^i \leftarrow w^i - \eta \frac{\partial \mathcal{L}(\mathbf{x}; w^i)}{\partial w^i}$
10:   **if** $s \sim \text{Bernoulli}(p) = 1$ **then**
11:     $\alpha^i \leftarrow \frac{\alpha^i}{2}$
12:     Send $\langle w^i, \alpha^i \rangle$ to randomly chosen worker.
13:   **end if**
14: **end loop**

Fig. 17. Program for the $i^{th}$ worker in GoSGD [45].

loss function landscape. As workers subsequently integrate results with their neighbors, which are in turn shared by these during the next iteration, eventually a remainder of each worker's findings travels in both directions around the ring. The more hops two workers are away from each other, the further they may diverge. Because the ring is closed, all workers project the same distance-attenuated force on each other which is crucial for stability [26].

*TreeEASGD* [25], which we illustrate along with the program that is run by each worker in Fig. 16, avoids bandwidth-related limitations by introducing a hierarchical tree-based communication pattern. Each worker implements two parameter exchange intervals. At every $\tau_{up}^i$ cycles, they share their current model parameters with the respective upstream node, and at every $\tau_{down}^i$ cycle, they share their current model parameters with all adjacent downstream nodes. Any received model parameters are immediately consumed by evaluating an $\ell^2$-norm-based function that penalizes the local model based on its divergence. The intuition here is that the intermediate workers explore the loss function landscape around the global model maintained by the root worker to locate descent trajectories that generalize well. Each leaf worker aids their parent intermediate worker in this endeavor by exploring closely related adjacent models. The degree of exploration is controlled by separately adjusting the up- and downstream parameter exchange frequencies (i. e. $\tau_{up}$ and $\tau_{down}$) for each worker based on its depth in the hierarchy.

Collaborative decentralized asynchronous training can also be realized without requiring the cluster to be structured explicitly. For example, all machines in *GoSGD* [45] are workers and can peer with each other to exchange parameters by implementing a sum-weighted gossip protocol [46].

Fig. 17 shows a pseudo-code implementation of the worker program. Each worker ($i$) defines a variable $\alpha^i$ that is initialized equally such that $\sum_i \alpha^i = 1$. During training, each worker repeatedly probes the loss function and updates its model. After each local update, a random Bernoulli distribution is sampled to decide whether a parameter exchange should be done. Thus, the probability ($p$) determines the average communication interval. If a parameter exchange is desired, a random node is chosen for peering. $\alpha^i$ is halved and sent along with the current model parameters ($w^i$) to the destination worker, which in turn replaces its local model with the weighted average based on its own and the received $\alpha$ value. Because the worker has absorbed this information, it adds the received $\alpha$ value to its respective value. Note that the symmetry holds, where $\sum_i \alpha^i = 1$, unless there is data in transit. Workers that shared their state recently are weighted down in relevance because the information they collected about the loss function has become more common knowledge (=gossip) among other workers, while those that accumulate changes without doing so gain weight. Thereby, the variance of staleness within the cluster is minimized. By setting the learning rate ($\eta$) to zero, all workers asymptotically approximate a consensus model.

## 4 EVALUATION

To gain an overview, we describe various existing DDLS and highlight their unique properties in Section 4.1. In Section 4.2, we align them in our taxonomy and discuss observations. In Section 4.3, we follow up our discussion by providing insights to aid choosing the right techniques.

### 4.1 Overview of Existing DDLS

Many works in distributed deep learning originate from the impressive results that were achieved in large-scale image classification at Google [18] using clusters consisting of tens of thousands of CPU nodes using *DistBelief*, a centralized asynchronous DDLS with support for multiple parameter servers [21]. Several DDLS have been proposed that extend the DistBelief approach. For example, *Petuum* [16] found that imposing delay bounds to control the staleness of asynchronous workers can improve the rate of convergence (cf. Section 3.3.4). *Parameter Server* [14] focused at formalizing the processing of deep learning workloads to establish

IEEE TRANSACTIONS ON PARALLEL AND DISTRIBUTED SYSTEMS, VOL. X, NO. X, XXXX                                                                                              13hybrid parallelism and integrate with a general machine learning architecture. *Project Adam* (Microsoft; [23]) took a similar approach by moving gradient computation steps into the parameter servers for some neural network layers. Additionally, they organize the parameter servers in a Paxos cluster to establish high availability. *TensorFlow* (Google; [17]) and *MXNet* (Apache Foundation; [19]) are modern descendants of DistBelief that improve upon previous approaches by introducing new concepts, such as defining backup workers (cf. Section 3.3.1), optimizing model partitioning using self-tuning heuristic models, and improving scalability by allowing hierarchical parameter servers to be configured (cf. Section 3.4.1).

Systems, such as *COTS HPC* [9] and *FireCaffe* [15], are reduced DDLS implementations that inherit certain ideas from DistBelief, but were optimized for HPC and GPU supercomputer environments, where they have been shown to achieve unparalleled performance for certain applications [18]. *CaffeOnSpark* (Yahoo; [24]) and *BigDL* (Intel; [27]) take the opposite approach and focus on easy integration with existing data analytics systems and commodity hardware environments by implementing centralized synchronous data-parallel model training on top of Apache Spark. To accommodate the frequent communication needs of such systems (cf. Section 3.3.1), they use sophisticated communication patterns to implement a distributed parameter server. The data-parallel optimizer of *PyTorch* (Facebook; [47]) takes a similar approach, but implements a custom interface to realize synchronous model training using collective communication primitives. Thereby, either one or all workers act as parameter server (all-reduce approach; cf. Section 3.4.1).

Most DDLS that rely on decentralized optimization stem from *SparkNet* [10], a decentralized synchronous DDLS that replicates Caffe-solvers using Apache Spark's map-reduce API to realize training in commodity cluster environments. As part of the popular Java DDLS *deeplearning4j*, this approach has gained widespread adoption. The restriction to synchronous execution is often considered as a major downside by this approach. Works that attempt to improve upon SparkNet are *MPCA-SGD* [11], which extends the basic Spark-based approach by overlapping computation and communication to realize quasi-asynchronous training and extrapolates the recently observed descent trajectory to cope with staleness, and *EASGD* [30], which retains the idea of limited isolated training phases but imposes fully asynchronous scheduling (Section 3.3.5). Having a single master node can become a bottleneck as the cluster grows larger. *D-PSGD* [26], *TreeEASGD* [25] and *GoSGD* [45] are approaches to further scale out decentralized optimization by distributing the master function (see also Section 3.4.2).

## 4.2 Taxonomic Perspective

In Table 1, we present a compact overview of all systems we have discussed. The taxonomy we developed in Section 3 serves as the basis of this overview.

**Parallelism**: DP is more frequently supported than MP, which can be explained as follows: 1) Decentral optimization is based on the concept of sparse communication between independent trainers. Realizing cross-machine MP in such systems is counter-intuitive; 2) New modeling and training techniques [2], [4] allow utilizing the available parameter space more efficiently, while technological improvements in hardware allow processing increasingly larger models. 3) Not every model can be partitioned evenly across a given number of machines, which leads to the underutilization of workers [20]. If a model fits well into the GPU memory, the resource requirements of the backpropagation algorithm can often be regulated reasonably well by adjusting the mini-batch size in DP systems. Therefore, some DDLS discourage using cross-machine MP in favor of DP, which is less susceptible to processing time variations [17].

**Optimization**: With respect the publication years, there is a trend towards decentralized systems in research. This can be explained with the rigid execution model of centralized systems (cf. Section 3.2.1). Recent contributions w.r.t. centralized DDLS can mostly be attributed to minor improvements, such as tailored optimization techniques [2], [36], [39] and the development of domain-specific compression methods [13]. However, centralized DDLS dominate industry usage and application research, although centralized and decentralized DDLS offer similar convergence guarantees [26]. A reason for this is certainly that centralized approaches are generally better understood and easier to use (cf. Section 4.3). Furthermore, most popular and industry-backed deep learning frameworks (PyTorch, TensorFlow, MXNet, etc.) contain centralized DDLS implementations that are mature, highly optimized and work tremendously well as long as parameter exchanges do not dominate the overall execution [11], [48].

**Scheduling**: Centralized asynchronous methods cope better with performance deviations and have the potential to yield a higher hardware utilization, but introduce new challenges such as concurrent updates and staleness (cf. Section 3.3.3), which is one reason why some DDLS support synchronous and asynchronous modes of operation [17], [19]. Note that centralized bounded asynchronous DDLS can always simulate synchronous and asynchronous scheduling (e. g. if a delay bound is used, $s = 0$ is identical to synchronous, while $s = \infty$ results in fully asynchronous behavior; cf. Section 3.3.4). Some decentralized DDLS define a simple threshold ($\tau$) to limit the amount of exploration per training phase (cf. Section 3.3.2). Others take a more dynamic approach to cope better with bandwidth limitations, which is indicated using the term *soft-bounded*.

**Parameter Exchange** mechanism: Binomial tree methods scale worse than scattering operations, but are preferable in high latency environments because less individual connections between nodes are required [44]. The listed, collective operation represents a common, but not necessarily the only parameter exchange method available. Some synchronous DDLS implement several collective operations and switch between them to maximize efficiency.

**Topology**: The current state-of-the-art in centralized DDLS for small clusters is the synchronous all-reduce-based approach (cf. Section 3.4.1). Large and heterogeneous setups can be utilized efficiently using hierarchically structured asynchronous communication patterns [19]. For decentralized systems, heavily structured communication protocols [25], [26], boosting techniques [11], as well as relatively unstructured methods [45] have been reported to offer better convergence rates than naïve implementations.

This is the author's version of an article that has been published in this journal. Changes were made to this version by the publisher prior to publication.
The final version of record is available at    http://dx.doi.org/10.1109/TPDS.2020.3003307

Copyright (c) 2020 IEEE. Personal use is permitted. For any other purposes, permission must be obtained from the IEEE by emailing pubs-permissions@ieee.org.

TABLE 1
Comparison and overview for different DDLS. We list *model / data parallelism* support as described in Section 3.1, support for centralized / decentralized *optimization* as described Section 3.2, scheduling method (cf. Section 3.3), the model *parameter exchange* mechanism and typical gathering/distribution method (data-parallel DDLS only), the overall *topology* employed to organize communication across the cluster (cf. Section 3.4), and *remarks* regarding interesting aspects or properties that may set the DDLS apart from other systems.

| DDLS Name (a-z) | Parallelism (Model / Data) | Optimi-zation | Scheduling | Parameter Exchange | Topology | Remarks |
|---|---|---|---|---|---|---|
| BigDL[27] | DP only | central | sync. | scatter-red. | distributed PS (always $k = n$) | Each worker acts as a parameter server for $\frac{1}{n}$ of the model (cf. Section 3.4.1). Distributed parameter exchanges are realized via the Spark block manager. |
| CaffeOnSpark[24] | DP only | central | sync. | scatter-red. | distributed PS (always $k = n$) | Parameter exchange realized via RDMA using repeated invocations of MPI functions. Equivalent implementations are available for **Caffe2** and **Chainer**. |
| COTS HPC[9] | MP only | central | sync. | – | distrib. array abstraction | Model layers partitioned along tensor dimensions and distributed across cluster. Fine-grained access is managed via a low-level array abstraction. |
| D-PSGD[26] | DP only | decentral | sync. | 2:1 reduce | closed ring | Each node exchanges parameters with only its neighbors on the ring (cf. Section 3.4.2). |
| DistBelief[21] | MP + DP | central | async. | ad hoc | distrib. PS | Model partitions spread across dedicated parameter server nodes (cf. Section 3.4.1). |
| EASGD[30] | DP only | decentral | async. | ad hoc | single master | Decentralized asynchronous system as discussed in Section 3.3.5. Reactive adjustment of hyper-parameters can speedup training [42]. |
| FireCaffe[15] | DP only | central | sync. | binom. tree | single PS | Simplistic centralized synchronous system as discussed in Section 3.3.1. |
| GoSGD[45] | DP only | decentral | soft-bounded async. | ad hoc | p2p mesh | No dedicated master node. Parameter exchanges between any two workers realized via sum-weighted randomized gossip protocol as discussed in Section 3.4.2. |
| MPCA-SGD[11] | DP only | decentral | soft-bounded async. | binom. tree | dedicated master node | Model updating and sharing updates are decoupled. Penalization occurs as a part of the model's cost function. Staleness effects are dampened using an extrapolation mechanism. |
| MXNet[19] | MP + DP | central | bounded async. | scatter-reduce *async.:* ad hoc | distributed PS (default $k = n$) | Supports various advanced parameter server configurations, including but not limited to hierarchical multi-stage proxy servers (cf. Section 3.4.1). |
| Parameter Server[14] | MP + DP | central | bounded async. | reduce *async.:* ad hoc | distrib. PS | Model partitions spread redundantly across parameter server group. Workers organized in model parallelism enabled groups. One worker per group can act as a proxy server. |
| Petuum[16] | MP + DP | central | bounded async. | ad hoc with eager scatter | distrib. PS | Pioneered the use of delay bounds to control staleness (cf. Section 3.3.4). Average model staleness is further reduced through the eager distribution of model parameters. |
| Project Adam[23] | MP + DP | central | async. | ad hoc | distrib. PS | Dedicated parameter server group that is managed as a Paxos cluster. Hybrid parallelism realized through transferring gradient computation for fully connected layers into PS. |
| PyTorch[47] | MP + DP | central | sync. | all-reduce | single PS or replicated PS | Model parallelism capabilities were added recently with version 1.4.0. Can only use either synchronous data-parallelism or model parallelism. |
| SparkNet[10] | DP only | decentral | sync. | reduce | dedicated master node | Decentralized synchronous implementation as discussed in Section 3.3.2. Realized using Spark map-reduce. Production-grade re-implementation present in **deeplearning4j**. |
| TensorFlow[17] | MP + DP | central | bounded async. | scatter/all-red. *async.:* ad hoc | distributed PS (default $k = n$) | Supports single and multi parameter server setups, as well as all-reduce-based approaches. By default, each worker acts as a parameter server for a portion of the model. |
| TreeEASGD[25] | DP only | decentral | bounded async. | ad hoc | tree | All nodes are workers and form a tree. Each worker only exchanges parameters with its immediate up- and downstream neighbors (cf. Section 3.4.2). |

## 4.3 Selecting the Right Technique

The non-linear non-convex nature of deep learning models in combination with the abundance of distributed methods opens up a large solution space [2]. When operating in distributed environments, each component can alter the optimal configuration, which makes direct comparisons difficult in practice. Although frequently done, comparing DDLS based on processing metrics such as GPU utilization or training sample throughput is not useful in practice, because such performance indicators can easily be maximized by increasing the batch-size, allowing more staleness (see Section 3.3.3) or extending exploration phases (see Section 3.3.2), which does not necessarily equate to faster training or yield a better model.

Well-established deep learning benchmarks like DAWN-Bench [48] propose comparing the end-to-end training performance by measuring quality metrics (e. g. time to accuracy x%). However, optimal configurations w.r.t. quality metrics are usually highly task dependent and may vary as the training progresses [42]. Performing exhaustive hyper-parameter searches is already a hard problem when training deep learning models using a single GPU [28]. The various side effects of the techniques we described in Section 3 further complicate the matter. Therefore, the collection and quantitative study of the performance of DDLS using standardized AI benchmarks is becoming increasingly important and can provide guidance regarding what configurations work well in practice. AI benchmarks differ in scope. DAWNBench [48], has a strong emphasis on distributed implementations, but focuses only on a few workloads.

TABLE 2
Criteria for selecting a DDLS techniques.

| | $RLBD_1$ | $RBM_2$ | $OptLR_3$ | $COCC_4$ | $CAHP_5$ | $RSOI_6$ | $ERB_7$ |
|---|---|---|---|---|---|---|---|
| **MP** | | ✓ | higher | + | +++ | + | trivial |
| **MP + mini-batch pipelining** | | ✓ | lower | ++ | ++++ | ++ | medium |
| **DP + central + synchronous** | ✓ | | higher | + | + | + | easy |
| **DP + central + asynchronous** | ✓ | | lower | +++ | ++ | +++ | hard |
| **DP + decentral + synchronous** | ✓ | | likely lower | + | ++ | ++ | easy |
| **DP + decentral + asynchronous** | ✓ | | | ++ | +++ | +++ | hard |

1 **R**equires **L**arge **B**alanced **D**ataset; 2 **R**equires **B**alanced **M**odel; 3 **Op**timal **L**earning **R**ate as cluster grows; 4 **C**omplexity due to **O**verlapping **C**omputation and **C**ommunication (e. g. growing staleness with cluster size); 5 **C**omplexity due to **A**dditional **H**yper-**P**arameters (number, stability, entanglement, etc.); 6 **R**esilience to **S**poradic **O**utside **I**nfluences; 7 difficulty to **E**stablish **R**eproducible **B**ehavior

Its successor MLPerf [49] expands the scope and defines stricter test protocols to establish better comparability. Other new benchmarks focus on gathering more information, for example, Deep500 [50] by defining metrics and measurement points along the training pipeline, or AIBench [51], which aims at covering many machine learning applications like recommendation systems, speech recognition, image generation, image compression, text-to-text translation, etc.

Aside from benchmark repositories, the criteria listed in Table 2 can help practitioners to narrow the scope when implementing or selecting a DDLS.

A first indicator is the dataset. Data-parallel training is most beneficial if the dataset itself is large enough, such that the additional concurrently sampled mini-batches increase variance. However, at the same time all data-parallel methods, except centralized synchronous approaches, require the mini-batches sampled by the workers to approximate the overall distribution$^{\triangleright\text{RLBD}}$. Model-parallel approaches do not have this restriction, but typically work best if the model can be partitioned evenly$^{\triangleright\text{RBM}}$.

Models for large-scale deep learning tasks are typically developed on multi-GPU workstations using scaled down datasets, and, once they deliver satisfactory results, they are then transferred to a cluster to train on *the complete dataset*. The optimal hyper-parameter setup can change in this process. Naturally, it is desirable to choose distributed training methods that do not further increase complexity. This is where simple model-parallel and synchronous data-parallel approaches outshine other methods, because the additional resources of the cluster are used to increase the effective mini-batch size. Thereby, the variance in the gradients decreases [31]. Thus, one can expect that compared with a single GPU implementation, the same or even a more aggressive learning rate can be used$^{\triangleright\text{OptLR}}$. The opposite is true for asynchronous and mini-batch-pipelining-based methods (cf. Section 3), where effects, such as increasing staleness$^{\triangleright\text{COCC}}$, need to be addressed by applying mitigation techniques (=adding complexity) and/or gradually decreasing the learning rate as the cluster size increases. The independent exploration principle employed in decentralized DDLS implies that the optimal local learning rates should decrease as well. However, hyper-parameters of the local optimizers and DDLS are often entangled$^{\triangleright\text{CAHP}}$, which usually makes costly full hyper-parameter swipes necessary to find good configurations [11].

Not all techniques cope equally well with instabilities in the cluster environment. The enforced coherence that allows choosing higher learning rates in model-parallel and synchronous data-parallel DDLS makes them more susceptible to outside influences$^{\triangleright\text{RSOI}}$, such as sporadic network bandwidth limitations due to concurrently running applications in the cluster (e.g. if some workers also act as Hadoop datanodes). Decentralized and/or asynchronous DP methods are significantly more resilient towards such situations and have been shown to vastly outperform other methods in bandwidth-constrained scenarios [8], [25].

Another aspect to consider is reproducibility, which is a general problem in deep learning, because the underlying numeric libraries often take different approaches to handle race conditions and rounding. The resilience to stragglers in asynchronous DDLS depends to a large degree on their ability to ignore race conditions. If reproducibility is desired, these methods often cannot function efficiently$^{\triangleright\text{ERB}}$.

## 5 REMARKS & FUTURE RESEARCH DIRECTIONS

In this survey, we discussed various theoretical and practical aspects that can arise when training deep learning models in a cluster, and offered suitable intuitions that allow reasoning about how different approaches utilize the available resources to realize collaboration. Thereby, we organized the underlying design principles that have a defining influence into a taxonomy. Our taxonomy establishes a basic schema that allows differentiating the major streams of distributed deep learning. We then applied this taxonomy to categorize a variety of DDLS independent of implementation-specific particularities to gain a compact overview.

Some possible areas of future research include: 1) Using decentralized optimization techniques in conjunction with P2P model sharing [45] could be an interesting area of research for certain IoT or automotive applications. 2) Most works in distributed deep learning restrict themselves to ideal test scenarios. In actual cluster setups the situation is usually more complex due to competing workloads. A comprehensive analysis of different distributed approaches in real-life scenarios would be helpful to many practitioners. Also, as an investment commodity, clusters are often not replaced, but rather extended [11]. Efficiently realizing distributed training in heterogeneous setups is a largely untackled engineering problem. 3) There exist various different benchmarking frameworks for deep learning workloads. A structured quantitative analysis of the results from these benchmarks could be interesting for many practitioners.

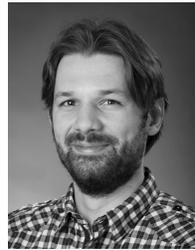

**Matthias Langer** received a PhD degree in computer science from LaTrobe University in 2019 and is currently employed at the Career Science Lab of BOSS ZhiPin in Beijing, where he engages in deep learning research to enable companies to find and recruit talented individuals in China and around the world. His research interests include large scale deep learning, high performance, and distributed computing.

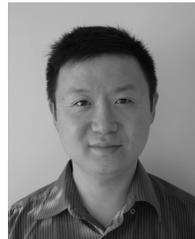

**Zhen He** is an associate professor in the Department of Computer Science at LaTrobe University. He leads a research group focused on applying deep learning to both image and text domains. His interests include distributed deep learning, human pose estimation, video activity recognition, using deep learning for dialogue systems, and application of deep learning to medical imagining data.

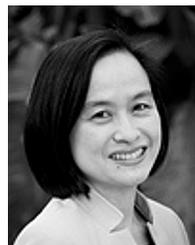

**Wenny Rahayu** is a professor in computer science and the Head of School of Engineering and Mathematical Sciences at LaTrobe University. The main focus of her research includes heterogeneous data integration, and mobile and distributed databases. In the last 10 years, she has published two authored books, three edited books and 200+ journal papers and conference proceedings, with more than 4000 total citations.

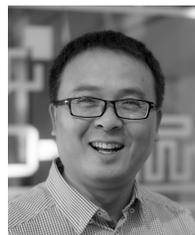

**Yanbo Xue** received his PhD degrees in electrical and computer engineering from McMaster University, and control theory and control engineering from Northeastern University (Shenyang, China). He is the Chief Scientist of Kanzhun Technology and the director of Career Science Lab in Beijing. His major research interests include deep learning, quantum computing, cognitive signal processing, and career science.